\documentclass[aip,rsi,superscriptaddress,graphicx,reprint]{revtex4-2}

\usepackage{graphicx}% Include figure files
\usepackage{dcolumn}% Align table columns on decimal point
\usepackage{bm}% bold math
\usepackage[utf8]{inputenc}
\usepackage[T1]{fontenc}
\usepackage{mathptmx}
\usepackage{etoolbox}
\usepackage{xcolor}
\begin{document}

% Use the \preprint command to place your local institutional report number
% on the title page in preprint mode.
% Multiple \preprint commands are allowed.
%\preprint{}

\title{The `n2EDM MSR' - a very large magnetically shielded room
with an exceptional performance
for fundamental physics measurements}

% repeat the \author .. \affiliation  etc. as needed
% \email, \thanks, \homepage, \altaffiliation all apply to the current author.
% Explanatory text should go in the []'s,
% actual e-mail address or url should go in the {}'s for \email and \homepage.
% Please use the appropriate macro for the type of information

% \affiliation command applies to all authors since the last \affiliation command.
% The \affiliation command should follow the other information.

\author{N.\,J.~Ayres}
\affiliation{ETH Z\"urich, Switzerland}

\author{G.~Ban}
\affiliation{Normandie Universit\'e, ENSICAEN, UNICAEN, CNRS/IN2P3, LPC Caen, Caen, France}

\author{G.~Bison}
\thanks{\bf Corresponding author, e-mail: georg.bison@psi.ch}
\affiliation{Paul Scherrer Institut, Villigen, Switzerland}

\author{K.~Bodek}
\affiliation{Jagiellonian University, Cracow, Poland}

%\author{N.~N.}
%\affiliation{Jagiellonian University, Cracow, Poland}

\author{V.~Bondar}
\affiliation{ETH Z\"urich, Switzerland}

\author{T.~Bouillaud}
\affiliation{Laboratoire de Physique Subatomique et de Cosmologie, Grenoble, France}

\author{B.~Clement}
\affiliation{Laboratoire de Physique Subatomique et de Cosmologie, Grenoble, France}

\author{E.~Chanel}
\altaffiliation[current address ]{Institut-Laue-Langevin, Grenoble, France}
\affiliation{Laboratory for High Energy Physics and
Albert Einstein Center for Fundamental Physics,
University of Bern, Bern, Switzerland}

\author{P.-J.~Chiu}
\altaffiliation[also at ]{ETH Z\"urich, Switzerland}
\affiliation{Paul Scherrer Institut, Villigen, Switzerland}

\author{C.\,B.~Crawford}
\affiliation{University of Kentucky, Lexington, United States of America}

%\author{N.~N.}
%\affiliation{University of Kentucky, Lexington, United States of America}

\author{M.~Daum}
\affiliation{Paul Scherrer Institut, Villigen, Switzerland}

\author{C.\,B.~Doorenbos}
\altaffiliation[also at ]{ETH Z\"urich, Switzerland}
\affiliation{Paul Scherrer Institut, Villigen, Switzerland}

\author{S.~Emmenegger}
\affiliation{ETH Z\"urich, Switzerland}

\author{A.~Fratangelo}
\affiliation{Laboratory for High Energy Physics and
Albert Einstein Center for Fundamental Physics,
University of Bern, Bern, Switzerland}

%\author{L.~Ferraris}
%\affiliation{Laboratoire de Physique Subatomique et de Cosmologie, Grenoble, France}

\author{M.~Fertl}
\affiliation{Institute of Physics, Johannes Gutenberg University, Mainz, Germany}

%\author{N.~N.}
%\affiliation{Institute of Physics, Johannes Gutenberg University, Mainz, Germany}

\author{W.\,C.~Griffith}
\affiliation{University of Sussex, Brighton, United Kingdom}

%\author{N.~N.}
%\affiliation{University of Sussex, Brighton, United Kingdom}

\author{Z.\,D.~Grujic}
\affiliation{Institute of Physics, Photonics Center, University of Belgrade, Serbia}

%\author{N.~N.}
%\affiliation{Institute of Physics, Photonics Center, University of Belgrade, Serbia}

\author{P.\,G.~Harris}
\affiliation{University of Sussex, Brighton, United Kingdom}

\author{K.~Kirch}
\affiliation{ETH Z\"urich, Switzerland}
\affiliation{Paul Scherrer Institut, Villigen, Switzerland}

%\author{N.~N.}
%\affiliation{ETH Z\"urich, Switzerland}
%\affiliation{Paul Scherrer Institut, Villigen, Switzerland}

\author{J.~Krempel}
\affiliation{ETH Z\"urich, Switzerland}

\author{B.~Lauss}
\thanks{\bf Corresponding author, e-mail: bernhard.lauss@psi.ch \\URL: www.psi.ch/en/ltp-ucn-physics}
\affiliation{Paul Scherrer Institut, Villigen, Switzerland}

\author{T.~Lefort}
\affiliation{Normandie Universit\'e, ENSICAEN, UNICAEN, CNRS/IN2P3, LPC Caen, Caen, France}

%\author{N.~N.}
%\affiliation{Laboratoire de Physique Corpusculaire, Caen, France}

\author{O.~Naviliat-Cuncic}
\affiliation{Normandie Universit\'e, ENSICAEN, UNICAEN, CNRS/IN2P3, LPC Caen, Caen, France}

\author{D.~Pais}
\affiliation{Paul Scherrer Institut, Villigen, Switzerland}

\author{F.\,M.~Piegsa}
\affiliation{Laboratory for High Energy Physics and
Albert Einstein Center for Fundamental Physics,
University of Bern, Bern, Switzerland}

%\author{N.~N.}
%\affiliation{Laboratory for High Energy Physics and
%Albert Einstein Center for Fundamental Physics,
%University of Bern, Bern, Switzerland}

\author{G.~Pignol}
\affiliation{Laboratoire de Physique Subatomique et de Cosmologie, Grenoble, France}

%\author{N.~N.}
%\affiliation{Laboratoire de Physique Subatomique et de Cosmologie, Grenoble, France}

\author{G.~Rauscher}
\affiliation{VAC - Vacuumschmelze, Hanau, Germany}

\author{D.~Rebreyend}
\affiliation{Laboratoire de Physique Subatomique et de Cosmologie, Grenoble, France}

\author{I.~Rien\"acker}
\altaffiliation[also at ]{ETH Z\"urich, Switzerland}
\affiliation{Paul Scherrer Institut, Villigen, Switzerland}

\author{D.~Ries}
\affiliation{Department of Chemistry – TRIGA site, Johannes Gutenberg University Mainz, Mainz, Germany}

%\author{N.~N.}
%\affiliation{Department of Chemistry – TRIGA site, Johannes Gutenberg University Mainz, Mainz, Germany}

\author{S.~Roccia}
\affiliation{Laboratoire de Physique Subatomique et de Cosmologie, Grenoble, France}

\author{D.~Rozpedzik}
\affiliation{Jagiellonian University, Cracow, Poland}

\author{W.~Saenz-Arevalo}
\affiliation{Normandie Universit\'e, ENSICAEN, UNICAEN, CNRS/IN2P3, LPC Caen, Caen, France}

\author{P.~Schmidt-Wellenburg}
\affiliation{Paul Scherrer Institut, Villigen, Switzerland}

\author{A.~Schnabel}
\affiliation{Physikalisch-Technische Bundesanstalt, Berlin, Germany}

%\author{N.~N.}
%\affiliation{Physikalisch Technische Bundesanstalt, Berlin, Germany}

\author{N.~Severijns}
%\affiliation{Katholieke Universiteit, Leuven, Belgium}
\affiliation{KU Leuven University, Belgium}

%\author{N.~N.}
%\affiliation{Katholieke Universiteit, Leuven, Belgium}

\author{B.~Shen}
\affiliation{Department of Chemistry – TRIGA site, Johannes Gutenberg University Mainz, Mainz, Germany}

\author{M.~Staab}
\affiliation{VAC - Vacuumschmelze, Hanau, Germany}

\author{K.~Svirina}
\affiliation{Laboratoire de Physique Subatomique et de Cosmologie, Grenoble, France}

\author{R.~Tavakoli Dinani}
%\affiliation{Katholieke Universiteit, Leuven, Belgium}
\affiliation{KU Leuven University, Belgium}

\author{J.~Thorne}
\affiliation{Laboratory for High Energy Physics and
Albert Einstein Center for Fundamental Physics,
University of Bern, Bern, Switzerland}

%\author{S.~Touati}
%\affiliation{Laboratoire de Physique Subatomique et de Cosmologie, Grenoble, France}

%\author{R.~Virot}
%\affiliation{Laboratoire de Physique Subatomique et de Cosmologie, Grenoble, France}
%

%\author{J.~Voigt}
%\affiliation{Physikalisch Technische Bundesanstalt, Berlin, Germany}
%

\author{N.~Yazdandoost}
\affiliation{Department of Chemistry – TRIGA site, Johannes Gutenberg University Mainz, Mainz, Germany}

\author{J.~Zejma}
\affiliation{Jagiellonian University, Cracow, Poland}

\author{G.~Zsigmond}
\affiliation{Paul Scherrer Institut, Villigen, Switzerland}
%

%\email[]{Your e-mail address}
%\homepage[]{Your web page}
%\thanks{}
%\altaffiliation{}

% Collaboration name, if desired (requires use of superscriptaddress option in \documentclass).
% \noaffiliation is required (may also be used with the \author command).
\collaboration{The nEDM collaboration}
%\noaffiliation

\date{\today}

\begin{abstract}
We present the magnetically shielded room (MSR) for the n2EDM experiment
at the Paul Scherrer Institute
which features
an interior cubic volume with
each side of length 2.92\,m,
thus providing an accessible space of
25\,m$^3$.
The MSR has
87 openings up to 220\,mm diameter to
operate the experimental apparatus inside,
and an intermediate space between the layers for sensitive
signal processing electronics.
The characterization measurements show a remanent magnetic field in the
central 1\,m$^3$ below 100\,pT, and a field below
600\,pT in the entire inner volume, up to 4\,cm to the walls.
The quasi-static shielding factor at 0.01\,Hz measured with a
sinusoidal 2\,$\mu$T peak-to-peak signal is about 100,000
in all three spatial directions and
rises fast with frequency to reach
10$^8$ above 1\,Hz.
\end{abstract}

%\pacs{3.14159}% insert suggested PACS numbers in braces on next line

\maketitle %\maketitle must follow title, authors, abstract and \pacs

% Body of paper goes here. Use proper sectioning commands.
% References should be done using the \cite, \ref, and \label commands

%\input{Introduction}

\section{Introduction}
\label{Introduction}

Magnetic shielding is used
when the absolute magnetic field strength
at a measurement site must be lower
than the Earth's magnetic field
or when Earth's or ambient magnetic field fluctuations
would limit the measurement accuracy.
%(There are many other applications as well.)

A common parameter to describe the performance of shields
is the shielding factor.
It is defined as the ratio of the 
magnetic flux density
%magnetic-field strength 
${\overrightarrow B}$
%$|B|$
measured at the center of the shield
and 
the magnetic flux density
%the field strength 
without shield
at the same position.

There are two classes of magnetic shields at room temperature,
active and passive, which can be used either
individually or in combination.
%,

Passive magnetic shields are built from high-permeability materials
which have a high `conductivity' for magnetic fields.
A shell of such material guides the external magnetic field
around an inner volume,
thus reducing the static magnetic field
as well as magnetic field variations in that volume.

The shielding effect of a passive shield of one layer is proportional
to the layer thickness.
For two separated layers the shielding
effect is the product of the shielding factors of the single shells
if the distance in between is large enough~\cite{Dubbers1986,VACBuch}.
Using multiple shield layers hence
reduces the required amount of the expensive high-permeability 
material necessary to achieve the same shielding factor,
but increases the volume of the shield walls.

The field guiding effect of high-permeability materials is the
dominating shielding effect only for magnetic disturbances with
frequencies below about 1\,Hz.
For these low frequencies the shielding factor approaches a constant value,
the quasi-static shielding factor,
% which is by convention
measured here with an excitation field oscillating at a frequency of $f_\mathrm{ex}=0.01$\,Hz.
%
%defined at 0.01\,Hz,
%which is the commonly used parameter
%to compare magnetic shields.
%
The strong increase of the shielding factor above 1\,Hz is caused
by the electrical conductivity of the shielding layer
and can be further increased by an additional `eddy-current' layer.
This is usually made of copper or copper-coated aluminum with a thickness of 5\,-\,12\,mm.

For magnetic field disturbances above 1\,kHz the shielding factor
is dominated by the radio-frequency (RF) shielding properties of the shield,
which would be perfect for an
%closed-conduction
electrically-closed conducting surface,
but in practice is limited by the size and design of the largest openings.
If the openings are designed as electrically conducting pipes
in the RF-shield,
the shielding effect can be maintained for larger frequencies
if the length to diameter ratio is
appropriately chosen.
%chosen according
%to the chimney dumping effect.
%
For most magnetic shields the incorporated eddy-current shield
is designed to simultaneously act as RF shield.

A static active shield uses a constant current in an
arrangement of coils to create a magnetic field
which compensates the surrounding field
in the volume of interest.
An dynamic active shield is a coil arrangement
additionally equipped with one or more
reference magnetic-field sensors
and a feedback control system
that adjusts the current source driving the coils
to compensate the detected magnetic-field
variations, see e.g.\ Refs.~\cite{Driscoll1971,Brake1991,Afach2014JAP,rawlik2018}.
The passive shield described in this article will be finally
surrounded by an active magnetic shielding installation
%for DC and AC fields
to further enhance the shielding performance
for frequencies below 5\,Hz.

A common passive shielding material is permalloy,
which is a nickel-iron alloy with a nickel
content above 75\%.
Various brand names with slightly different material compositions and properties exist.
%we use MUMETALL$^{\tiny \textregistered}$
%and ULTRAVAC$^{\tiny \textregistered}$.
%
%The shield presented in this work is based on different variants
%of MUMETALL\texttrademark\
%from VAC - Vacuumschmelze, Hanau\footnote{VACUUMSCHMELZE GmbH \& Co. KG, Gruener Weg 37, D-63450 Hanau, Germany}.
%
The high permeability is obtained by a special annealing process in a
reductive atmosphere
at temperatures above $\sim$1050\,$^{\circ}$C.
Another relevant manufacturing
factor
is the necessary careful handling of the material after annealing.
Mechanical stress on the material,
for example during bending,
reduces the permeability.
Large shields have to be assembled
from flat sheets and
edge pieces bent before annealing.

The first magnetic shields large enough for humans, used
to measure the magnetic field of the heart or brain, were
built in the 1960s~\cite{Cohen1967,Cohen1967b,Cohen1968,Cohen1970}.
Such large shields with two or more magnetic shielding layers
and door access,
called magnetically shielded rooms (MSR),
are nowadays commercially available from different companies.

Initially, the driving force for installation of large MSRs
with more than two layers~\cite{BMSR,helsinki,cosmos}
%%% ,Cohen2002}
were precise measurements of bio-magnetic fields in humans.
For many years the MSR with the highest shielding factor was `BMSR-2'
at the Physikalisch-Technische Bundesanstalt, Berlin, Germany
with originally seven, now eight, magnetic shielding layers~\cite{BMSR2}.
%
%%%%%%%%%%%%%

Large multilayer shields were pioneered also in
fundamental physics measurements
already in the 1980s, e.g.\ Refs.~\cite{Loving1977,Altarev1981PLB,Pendlebury1984,Lamoreaux1987,Smith1990PLB}.
One of the first large MSRs dedicated to physics experiments
was built at Oak Ridge National Laboratory~\cite{Soltner2011}
followed by one at Technical University Munich~\cite{Altarev2014RSI,Altarev2015}.

The MSR described in this work
serves the n2EDM experiment,
aiming at an improved measurement of the neutron electric dipole
moment (nEDM)~\cite{Ayres2021n2EDM}.
The key requirement for n2EDM, besides a high shielding factor, is the ability to
generate a very uniform magnetic field in the central 1\,m$^3$ volume of
the MSR~\cite{Abel2019}.
The MSR design has to meet those requirements while complying with mechanical
boundary conditions such as shield geometry, size, weight, number and size
of openings and accessibility.
Factors that affect the field uniformity are the magnetization state
of the shielding metal,
%the coils that generate the internal magnetic field
the homogeneity of a desired field produced by
an internal coil system
necessary for the nEDM experiment,
and disturbances caused by the openings.
Since those effects drop off with distance,
MSRs with a large inner volume facilitate a good magnetic field uniformity,
while MSRs with a smaller inner volume make it easier to achieve
large shielding factors~\cite{VACBuch}.
The design presented here is a compromise between these
two factors, which optimizes the overall performance of
our experiment.
In this publication we demonstrate that the realized design achieves both
a high shielding factor and a 
low and homogeneous enough magnetic field,
which results from low disturbances of field uniformity
due to the MSR's magnetization state.

%%%%%%%%%%%%%%%%%%%%%%%%%%%%%%%%%%%%%%%%%%%%%%%%%%%%%%%%%%%%%%%%%%%%%%%%%%%%%%%%%%%%%

\section{Design guidelines}
\label{Sec:Specification}

The design of the MSR was driven by
the performance needed to reach the sensitivity goals
of the n2EDM experiment
and by the
restrictions imposed by the
apparatus to be installed inside the MSR~\cite{Ayres2021n2EDM}.
Further constraints were set by the spatial dimensions of the
installation area within the experimental hall.
%and the available height for installation and crane maneuvers.

The number and the dimensions of the openings were
given by the components of the apparatus.
Two very large openings with a diameter of 220\,mm
are required for the
installation of the ultracold neutron guides.
As a design principle all openings are symmetrically mirrored on opposite MSR
walls, which helps to suppress first-order gradients.

The doors must provide
a minimum 2\,m$\times$2\,m
access for the inner chamber
to allow for equipment installation,
the largest one being the vacuum tank.

The key specified design performance criteria were:
1)
a quasi-static magnetic shielding factor
at 0.01\,Hz
of 70'000, and
2) a remanent magnetic field in the
central 1\,m$^3$ below 500\,pT
with a field gradient lower than
300\,pT/m.
%
%The construction an on-site assembly
%of the MSR was done by the company
%VAC - Vacuumschmelze
%\footnote{VACUUMSCHMELZE GmbH \& Co. KG, Gruener Weg 37, D-63450 Hanau, Germany}
%
%following the design together with the nEDM collaboration.

%%%%%%%%%%%%%%%%%%%%%%%%%%%%%%%%%%%%%%%%%%%%%%%%%%%%%%%%%%%%%%%%%%%%%%%%%%%%%%%%%%%%%

\section{Construction of the MSR}
\label{MSR-setup}

\begin{figure}[h]
\centering
\includegraphics[scale=0.5]{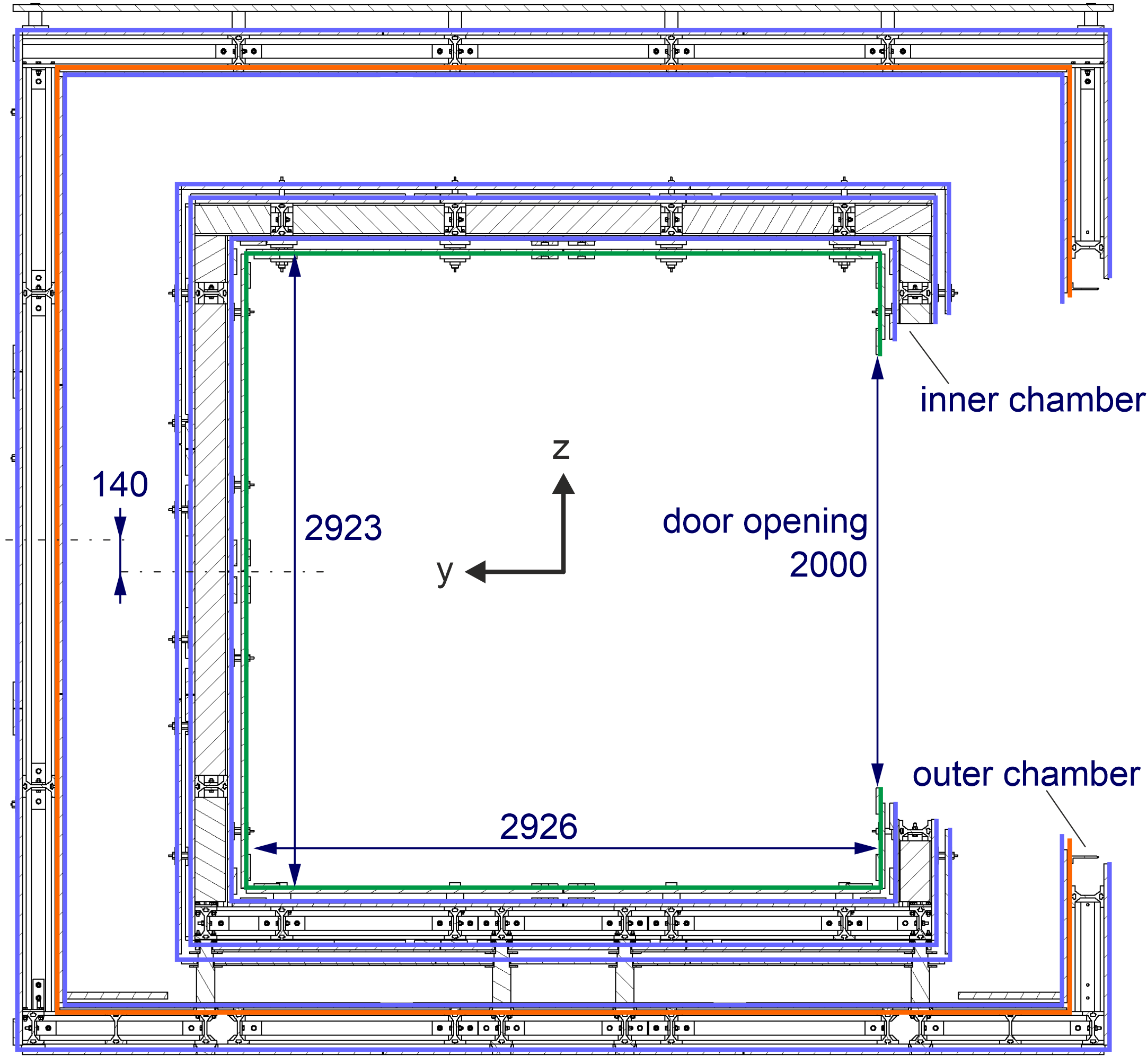}
\caption[CAD]{
Vertical cut showing the positioning and dimensions of the
inner and outer chambers with shielding layers,
with MUMETALL$^{{\tiny \textregistered}}$\ layers indicated in blue,
ULTRAVAC$^{{\tiny \textregistered}}$\ layer in green,
and
aluminum layer in red, as listed in Tab.~\ref{tab:layers}.
All dimensions are in mm.}
\label{fig:cut-view}
\end{figure}

The MSR was engineered, designed and constructed 
by VAC,
Germany\footnote{VACUUMSCHMELZE GmbH \& Co. KG, Gruener Weg 37, D-63450 Hanau, Germany}
in an iterative process with input from the nEDM collaboration.
%Following the design together with the nEDM collaboration,
%The MSR was constructed and built by VAC.
%Germany\footnote{VACUUMSCHMELZE GmbH \& Co. KG, Gruener Weg 37, D-63450 Hanau, Germany}.
%at the Hanau factory of VAC.
%
The high permeability materials used in all shielding
layers were produced via smelting from the original ores
in the furnaces of the VAC Hanau facility.

\begin{figure}[h]
\centering
\includegraphics[width=\columnwidth]{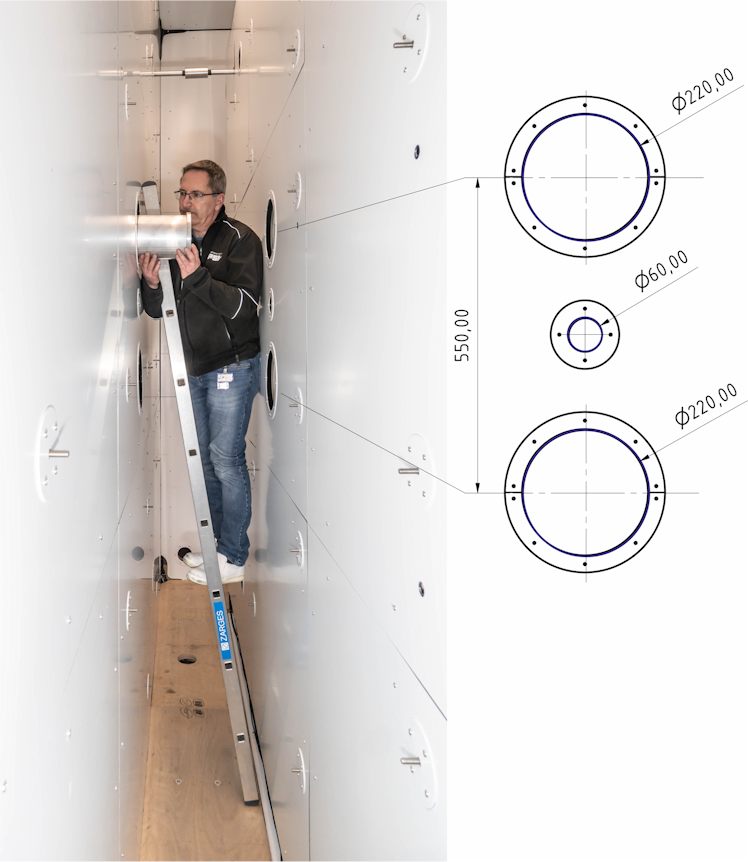}
\caption[]{
Intermediate space between inner and outer
chamber.
The photo shows a test with a vacuum tube
passing
through
one large opening.
The diagram to the right depicts the
dimensions and separation of the
large openings in the center of the wall.
All shown dimensions are in mm.
}
\label{fig:MSR-intermediate-space}
\end{figure}

The MSR design consists of an outer and an inner chamber,
and an intermediate space as shown in Fig.~\ref{fig:cut-view}.
The inner chamber is centrally placed
and separated from the outer chamber
horizontally by a distance of
approximately\ 45\,cm in all directions from the outer wall,
with a vertical offset of
14\,cm towards the floor.
The intermediate space between the chambers
shown in Fig.~\ref{fig:MSR-intermediate-space}
provides an area which is RF shielded,
and magnetically shielded with a quasi-static shielding factor of about 65.
It is accessible for persons and
can be used for additional experimental equipment
as well as sensitive signal electronics
that are too magnetic to be located
next to the central n2EDM apparatus.
%inside the MSR.
%
The outside dimensions of the MSR are 5.2\,m$\times$5.2\,m horizontally,
and 4.8\,m vertically.
The inner chamber is almost perfectly cubic with
a side length
of 2.92\,m,
thus featuring 25\,m$^3$ of internal volume for the installation of the experimental apparatus.
%
%A section view of the chamber assembly is shown in Fig.~\ref{fig:cut-view}.
%

\begin{figure}[h]
\centering
\includegraphics[width=\columnwidth]{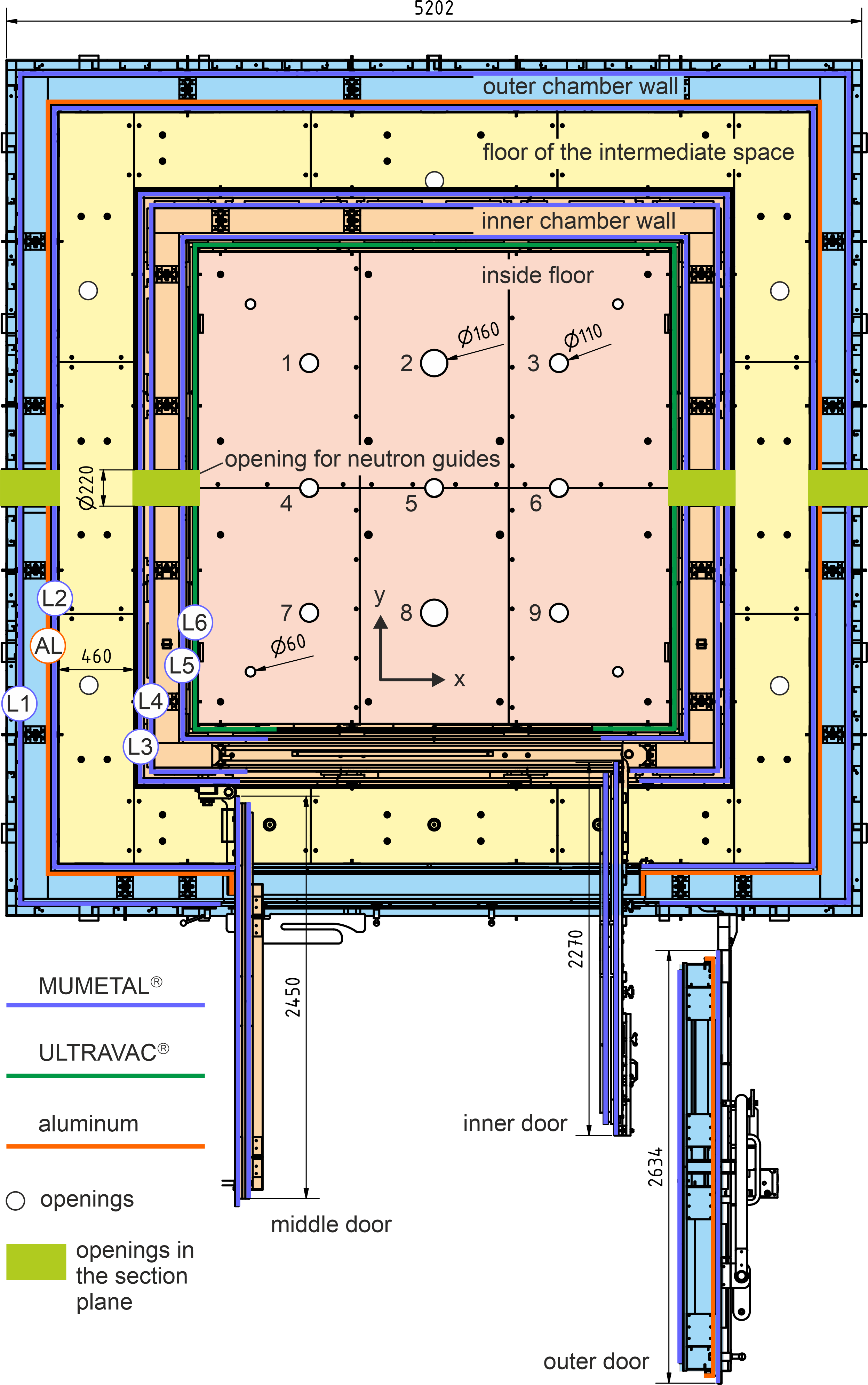}
\caption{
Vertical section view
onto the floor of the MSR.
All dimensions are in mm.
The layers are as given in
Tab.~\ref{tab:layers}.
There are two openings with an ID of 160\,mm, seven openings
with an ID of 110\,mm, and four openings with
an ID of 60\,mm in the
central region of the
floor (roof).
The pattern of openings in the floor is mirrored on the ceiling of the MSR.
Some openings are numbered to allow the identification of
measurement locations.
Additional openings in the outside wall
allow for external connections of the equipment that will be installed in the
intermediate space.
}
\label{fig:openings-floor}
\end{figure}

Many openings in all six walls of the MSR provide
access to and allow operation of the n2EDM apparatus
on the inside.
The two neutron guides require
the largest openings
with 220\,mm diameter, with their centers
separated by 550\,mm.
The diagram in Fig.~\ref{fig:MSR-intermediate-space} 
illustrates this arrangement.
Identical openings on opposite sides
of the chamber will be used for two pumping lines.
Furthermore, nine large openings are
symmetrically placed on the roof and the floor
(Fig.~\ref{fig:openings-floor}),
which will
be used for e.g. laser paths,
optical fibers,
cables and sensor tubing.
A few openings are only either in the inner or outer chamber.
%\newline
%
The total number of openings amounts to 87, planned with contingency:
\begin{itemize}
\item 4 with inner diameter (ID) = 220\,mm, %ok
\item 4 with ID = 160\,mm,
\item  43 with ID = 110\,mm (21 only in outer chamber),
\item 2 with ID = 80\,mm,
\item
26 with ID = 60\,mm (8 only in inner chamber),
%and
\item 8 with ID = 55\,mm.
\end{itemize}

Apart from the openings which are in one chamber only,
all openings are
coaxially passing
%in an aligned passage from the inside to the outside
through the inner and outer chamber walls.
%
%An impression of the large number of openings
%is provided in
Figure~\ref{fig:openings-floor}
provides a sense of the arrangement of the
openings in the floor of the MSR.

\begin{figure}[h]
\centering
\includegraphics[width=\columnwidth]{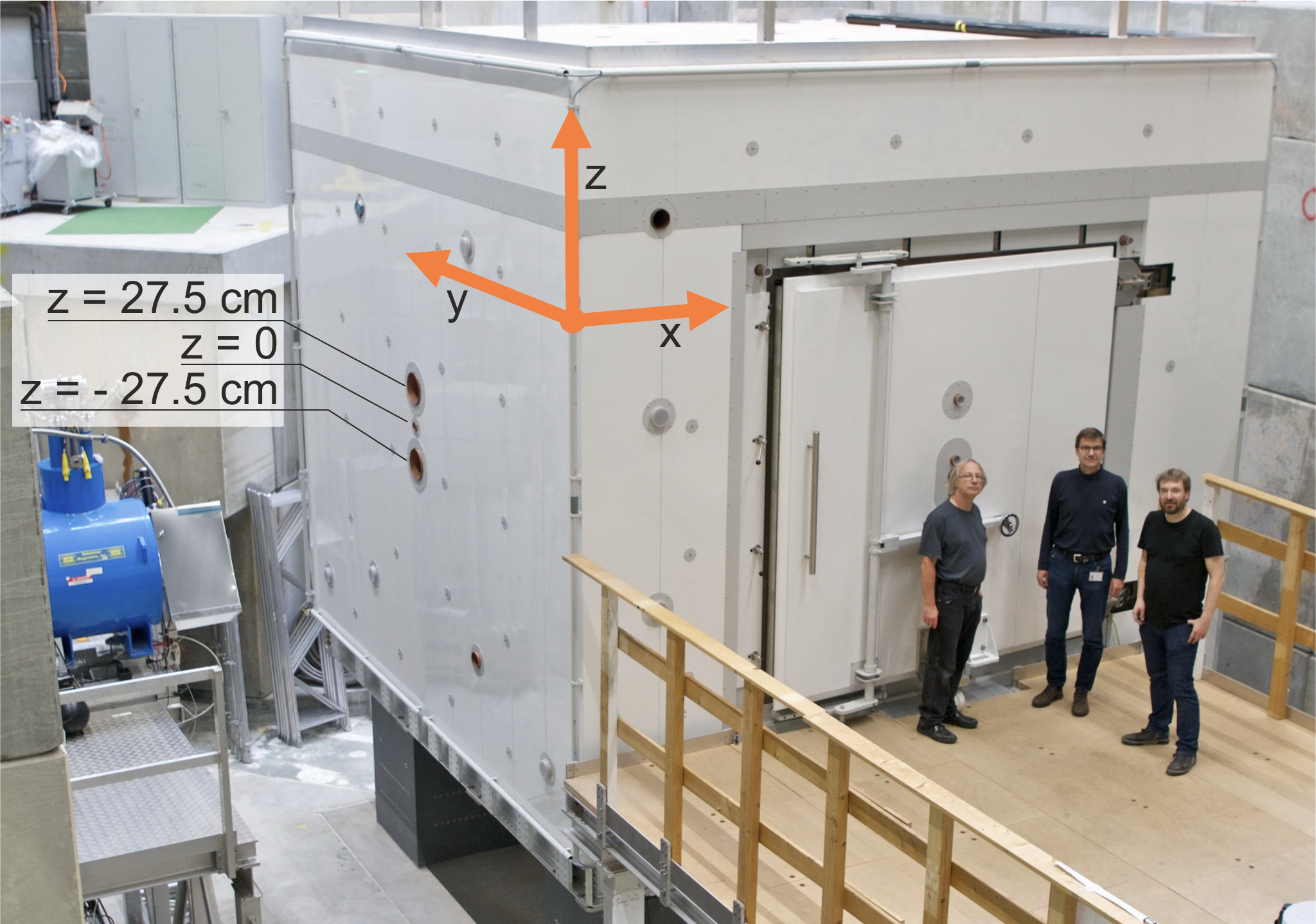}
\caption[]{
The MSR installed in area South of the PSI UCN source.
The experiment's coordinate system is
indicated.
The openings used for the horizontal scans along the x-axis (see Fig.~\ref{fig:Bz}) are labeled with their z-coordinates.
}
\label{fig:MSR}
\end{figure}

\begin{figure}[h]
\centering
\includegraphics[width=\columnwidth]{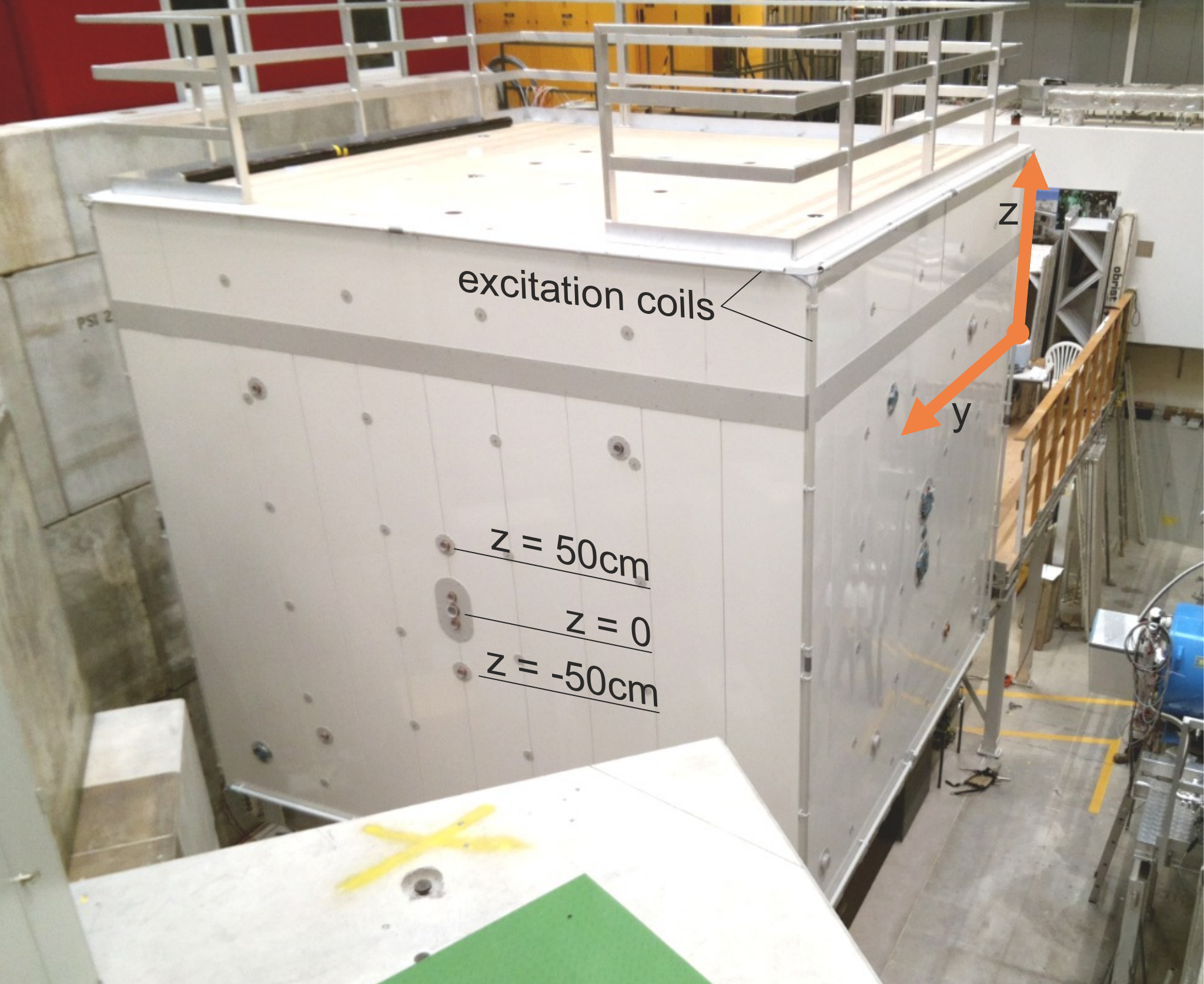}
\caption[]{
View of the rear side of the MSR.
Two of the tubes carrying the excitation coils used to measure the shielding factor are indicated.
The openings used for the horizontal scans along the y-axis (see Fig.~\ref{fig:Bxv}) are labelled with their z-coordinates. 
}
\label{fig:MSR-back}
\end{figure}

The assembled MSR,
installed in experimental area South
of the ultracold neutron (UCN) source~\cite{Bison2020,Lauss2021} at
the Paul Scherrer Institute (PSI),
is shown in Fig.~\ref{fig:MSR} from the side
of the entrance door and in
Fig.~\ref{fig:MSR-back} from the rear.
It is placed on an aluminum frame positioned on
four 1364\,mm high granite pillars with 1\,m $\times$1\,m base,
all placed
on top of its own
concrete foundation,
vibrationally isolated from the surrounding concrete floor
of the experimental hall.
%1364 wurde bestellt / 1365 sind von Bernhard schlecht gemessenm

\begin{figure}[h]
\centering
\includegraphics[width=\columnwidth]{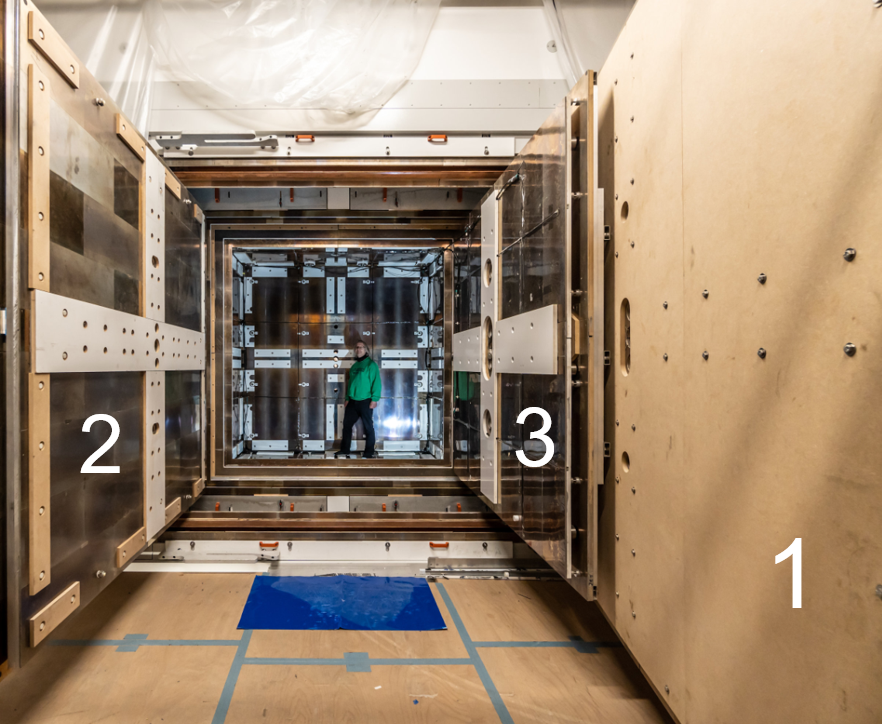}
\caption[]{
View into the MSR with
all three numbered doors open.
}
\label{fig:MSR-open}
\end{figure}

%\clearpage

\begin{table}[h]
\centering
\begin{tabular}{|l|c|c|c|}
    \hline
Chamber & Layer  &  Thickness (mm) &  Material \\ \hline
Outer  &   L1   &   3.75   &  MUMETALL$^{\tiny \textregistered}$  \\ \hline
Outer  &   Al  &   8.00   &  Aluminum   \\ \hline
Outer  &   L2   &   3.75   &  MUMETALL$^{\tiny \textregistered}$   \\ \hline
Inner  &   L3   &   6.75   &  MUMETALL$^{\tiny \textregistered}$   \\ \hline
Inner  &   L4   &   6.75   &  MUMETALL$^{\tiny \textregistered}$   \\ \hline
Inner  &   L5   &    4.5   &  MUMETALL$^{\tiny \textregistered}$   \\ \hline
Inner  &   L6   &    6.0   &  ULTRAVAC$^{\tiny \textregistered}$   \\ \hline
\end{tabular}
\caption{Naming scheme and thicknesses of all shielding layers.}
\label{tab:layers}
\end{table}

%The MSR has
%six independent magnetic shielding layers
%as detailed in table~\ref{tab:layers}
%%numbered from the outside to the inside
%and one eddy current shield.
%

The MSR consists of seven shielding layers
(Tab.~\ref{tab:layers}),
with one aluminum layer acting as eddy-current and RF shield.
Of the six soft magnetic layers,
the five outer ones are made of MUMETALL$^{\tiny \textregistered}$
(Ni 77\%, Cu 4.5\%, Mo 3.3\%, Fe balance),
a soft magnetic NiFe alloy with a Z-shaped hysteresis curve~\cite{Demper2015} %loop form
and correspondingly high maximum permeability.
MUMETALL$^{\tiny \textregistered}$ is a standard alloy for magnetic shielding.
However, the alloy ULTRAVAC$^{\tiny \textregistered}$816
(Ni 81\%, Mo 6\%, Fe balance)
employed for the innermost layer
was applied here for the first time in MSRs.
This novel NiFe alloy has a round-loop-shaped hysteresis curve
due to its composition~\cite{Demper2015}.
In this alloy, remagnetization processes take place mainly via reversible
domain wall motions.
This material is characterized by a high initial permeability even at
saturation levels of magnetic field strength H < 0.1\,A/m in the shielding layer,
and by a lower maximum permeability compared to MUMETALL$^{\tiny \textregistered}$.
Due to the round-loop-shaped hysteresis curve,
the remanence
of ULTRAVAC$^{\tiny \textregistered}$816
with a residual magnetic flux density B$_r$ = 0.2\,--\,0.3\,T is less than
half that of MUMETALL$^{\tiny \textregistered}$ (B$_r$ = 0.45\,--\,0.55\,T).
This hysteresis shape allows
%forms an ideal prerequisite
for an optimal
demagnetization of the innermost layer to achieve minimum residual fields.
All walls were manufactured
using the VAC proprietary panel technique.

All additionally materials used in the MSR construction
were previously checked
for magnetic contaminations with different specifications.
The most stringent restrictions
applied to materials in the inner chamber,
i.e. allowing for a maximum 200\,pT signal at 50\,mm distance,
when scanned in the BMSR-2 magnetic testing facility at PTB Berlin~\cite{BMSR2}.
%
%Styrofoam$^{{\tiny \textregistered}}$\
Expanded polystyrene
placed between the individual layers serves as thermal insulation.

%\clearpage

%Figure~\ref{fig:openings-floor} also shows the MSR with its three
%opened doors and gives the relevant dimensions.
%
Figure~\ref{fig:MSR-open} shows
a photograph of
the open MSR with the doors visible
on the sides.
Information about the dimensions can be found
in Fig.~\ref{fig:openings-floor}.
%
%The closed doors provide excellent contact between
%the MUMETALL$^{{\tiny \textregistered}}$ sheets of the door and the fixed walls.
All doors are larger than the door openings.
The overlap is necessary to reduce the magnetic resistance
for the field when passing from the wall to the door.
On all doors, dedicated aluminum plates allow the
mechanical contact pressure to be increased.
The opening and closing operations are fully manual and
can be done in about 20\,minutes.
With a weight of 1500\,kg, the outermost door needs to be supported by an additional wheel.

%%%%%%%%%%%%%%%%%%%%%%%%%%%%%%%%%%%%%%%%%%%%%%%%%%%%%%%%%%%%%%%%%%%%%%%%%%%%%%%%%%%%%

\section{Shielding factor}
\label{Shieldingfactor}

The shielding factor was measured using
excitation coils on the outside
edges of all outermost walls of the MSR (see Fig.~\ref{fig:MSR-back}).
%
%These coils were calibrated
%with a mapped magnetic field
%generated with an external
%additional coil system mounted
%before the MSR installation
%on a larger frame at a distance of about 1.5\,m
%to the MSR walls.
%
The coil constants $K_\mathrm{ex}$ of those
coils were calibrated with an additional
external coil system, which had been mounted
on a large frame before the installation
of the MSR.
The distance of these coils to the later
position of the MSR walls was approximately 1.5\,m.
The excitation coils produced a sinusoidal signal
$B_\mathrm{ex} = K_\mathrm{ex}\, I_\mathrm{ex}\, \sin(2 \pi\, f_\mathrm{ex}\, t)$
with 2\,$\mu$T peak-to-peak amplitude
at the MSR center position.
A QuSpin$^{\tiny \textregistered}$ magnetometer\footnote{
QuSpin, Inc.
331 South 104th Street, Suite 130
Louisville, CO 80027, USA}
recorded the excitation
signal inside the inner chamber.
The sensor was installed in the
center of the chamber, inside a small calibration coil
that generated a sinusoidal reference signal
$B_\mathrm{ref} = K_\mathrm{ref}\,I_\mathrm{ref}\, \sin(2 \pi\, f_\mathrm{ref}\, t) $
with the reference frequency $f_\mathrm{ref}$
well separated from the excitation frequency $f_\mathrm{ex}$.
The coil constant $K_\mathrm{ref}$ of the
reference coil was independently measured
and agrees to better then 1\% with the calculated value.
During data taking the magnetometer signal as well as
the monitor signals for the two currents in the coils $I_\mathrm{ex}$ and
$I_\mathrm{ref}$ were recorded by a multichannel
ADC which was synchronized with the function generator
that supplied the $f_\mathrm{ex}$ and $f_\mathrm{ref}$ signals.
The duration of the times series recorded
by the ADC for each test frequency
was programmed such that it contained an
exact integer multiple of the
oscillation periods of $f_\mathrm{ex}$
and $f_\mathrm{ref}$.
This simplified the data analysis since
each oscillation signal was guaranteed
to contribute only to a single frequency bin in
the Fast Fourier Transformation (FFT) spectrum of
the time series.
This method minimizes the influence of external
disturbances on the final result because noise
in frequency bins other than the ones centered
at $f_\mathrm{ex}$ and $f_\mathrm{ref}$
is disregarded.
The applied FFT algorithm extracted the
root-mean-square amplitudes of the signals at
the relevant frequencies.
Those were
the amplitude of the current in the excitation
coil $I_\mathrm{ex}^\mathrm{RMS}$,
the amplitude of the current in the reference
coil $I_\mathrm{ref}^\mathrm{RMS}$,
the magnetometer signal at the excitation
frequency $B_\mathrm{ex}^\mathrm{RMS}$,
and the magnetometer signal at the reference
frequency $B_\mathrm{ref}^\mathrm{RMS}$.
Comparing the measured reference signal to the expected
amplitude gave us an in-place correction factor
$C_\mathrm{cal}$ for the calibration of the magnetometer:
\begin{equation}
 C_\mathrm{cal} = \frac{B_\mathrm{ref}^\mathrm{RMS}}{K_\mathrm{ref}\,I_\mathrm{ref}^\mathrm{RMS}}
\end{equation}
The shielding factor $F_S$ results in a similar way
from comparing the measured amplitude at the excitation
frequency to the value calculated from the coil constant
and current:
\begin{equation}
F_S = \frac{K_\mathrm{ex}\,I_\mathrm{ex}^\mathrm{RMS}}{B_\mathrm{ex}^\mathrm{RMS}} \, C_\mathrm{cal}
= \frac{K_\mathrm{ex}\,I_\mathrm{ex}^\mathrm{RMS}}{B_\mathrm{ex}^\mathrm{RMS}} \, \frac{B_\mathrm{ref}^\mathrm{RMS}}{K_\mathrm{ref}\,I_\mathrm{ref}^\mathrm{RMS}}
\end{equation}
The measurement method makes the result independent of the
magnetometer calibration and depends only on amplitude
measurements and coil-constants which were independently
cross-checked.
The measurement was performed for the three spatial directions in almost the same way.
Only the density of excitation frequencies $f_\mathrm{ex}$ was increased for the $x$ and $z$ direction in order to investigate the noise above 5\,Hz.

\begin{figure}[t]
\centering
\includegraphics[width=\columnwidth]{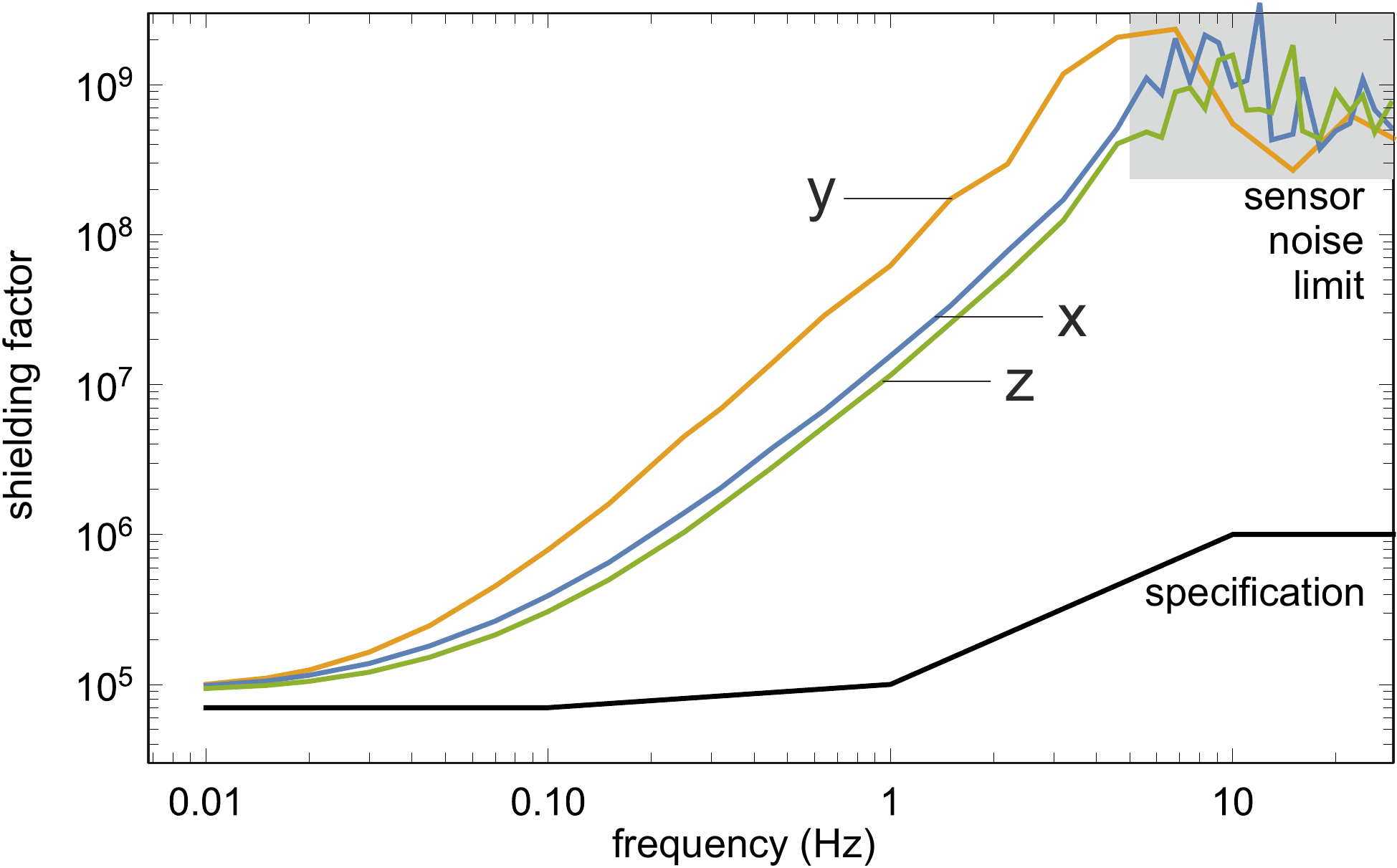}
\caption[]{
Dependence of the magnetic shielding factor on frequency measured
with a sinusoidal 2\,$\mu$T peak-to-peak
signal for the 3 spatial dimensions
as defined in Fig.~\ref{fig:MSR}.
The black line shows the specified minimum
required shielding factor for the depicted frequency range.
The gray shaded area shows the region where
due to the shield
the excitation signal is reduced to the level of the sensor noise.
}
\label{fig:shielding-factor}
\end{figure}

The measured frequency-dependent shielding factor is
shown in Fig.~\ref{fig:shielding-factor}.
At frequencies above 5\,Hz the shielding factor is so large
that the sensor reaches its noise limit.
%
%This noise limit itself is a function of frequency
%since the magnetic interference at PSI
%is not uniform in frequency
%and not stable in time.
%
Additionally, the measurement above 5\,Hz
shows interference from the PSI magnetic environment
which lead to fluctuating results,
with a minimum shielding factor of 10$^8$.
The specified performance is surpassed at all measured frequencies.

The quasi-static shielding factor at 0.01\,Hz,
which is
most important for the n2EDM experiment,
is $\sim$100,000 in all spatial directions,
namely 101,300$\pm$500 in $x$-direction,
101,000$\pm$1000 in $y$-direction,
and 94,900$\pm$1400 in $z$-direction.

Figure~\ref{fig:shielding-factor} also shows that
for frequencies below 5\,Hz the shielding factor in the $y$-direction
is consistently larger than in the other directions.
%
%An explanation is that the current
%induced by field fluctuations
%around the chamber in this direction is not crossing the door side,
%which has a lower electric conductivity
%due to the imperfect door contact,
%and the magnetic field lines can avoid the door overlaps
%by directly entering into the four walls around the y-direction.
%
This behavior is expected since the current
induced by the y-excitation coil is not 
crossing the door side,
which has a lower electric conductivity
due to the imperfect door contact,
and the magnetic field lines can avoid the door overlaps
by directly entering into the four walls around the y-direction.
A comparison of the performance in
the $x$- and $y$-directions thus
gives an estimate on the losses caused by the imperfect magnetic and
electric contacts of the doors.

\begin{figure}[h]
\centering
\includegraphics[scale=0.30]{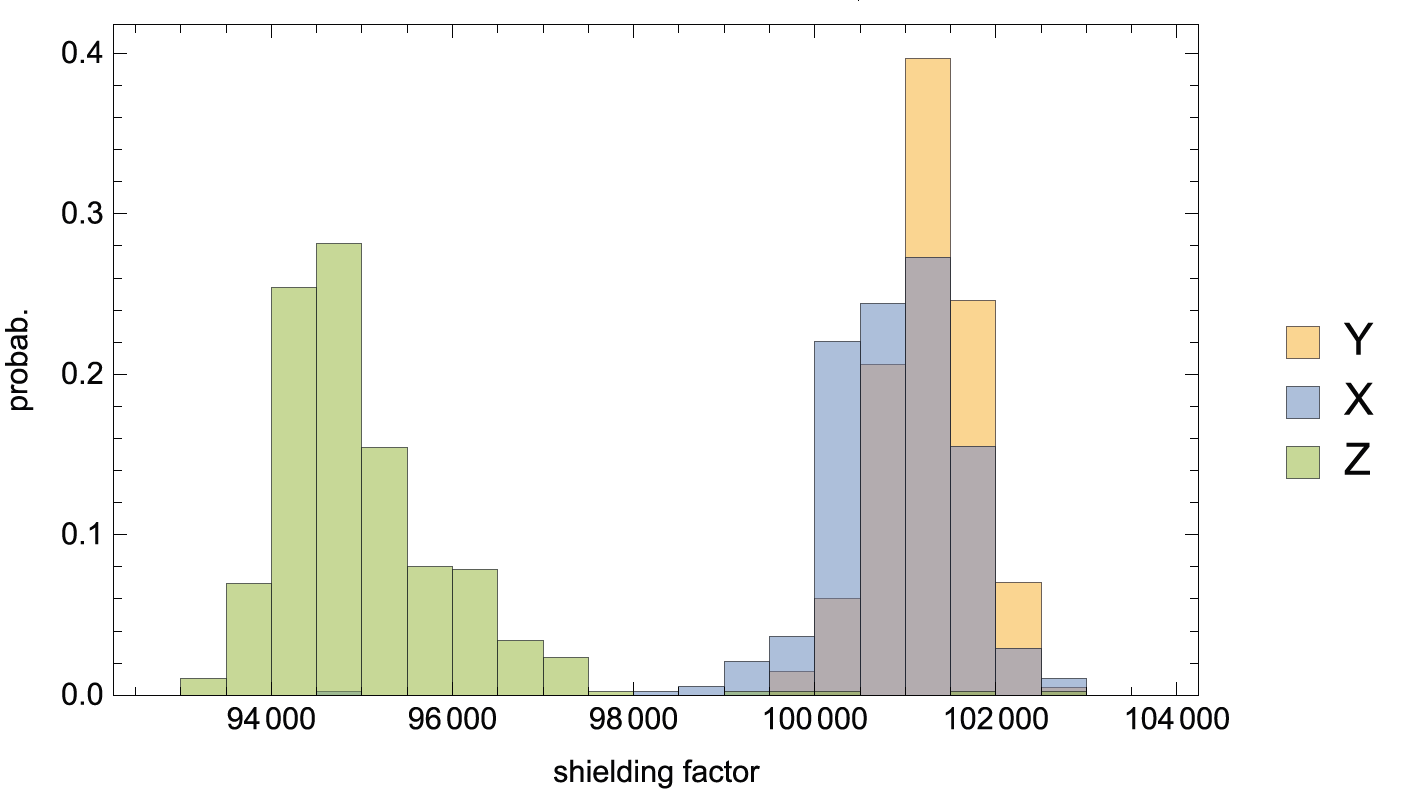}
\caption[]{
Histogram of all
measurements of the quasi-static shielding factor
at 0.01\,Hz in the three spatial directions.
}
\label{fig:shielding-factor-xyz}
\end{figure}

A histogram of the individual shielding factor
measurements at 0.01\,Hz is shown
in Fig.~\ref{fig:shielding-factor-xyz}.
The quasi-static shielding factor in the z-direction is slightly smaller
than in the other directions.
This is likely caused by the smaller distance 
and the offset
between the inner and outer chamber
in the vertical direction.
% which causes a stronger magnetic coupling in this direction.
The spread is likely due to a combination of the
statistical uncertainty of the measurement and the changing
magnetic environments over the course of the
measurements which also causes a small change of the shield response.
%

%%%%%%%%%%%%%%%%%%%%%%%%%%%%%%%%%%%%%%%%%%%%%%%%%%%%%%%%%%%%%%%%%%%%%%%%%

\section{Equilibration of MSR layers}
\label{Equilibration}

\begin{figure}[h]
\centering
\includegraphics[width=0.8\columnwidth]{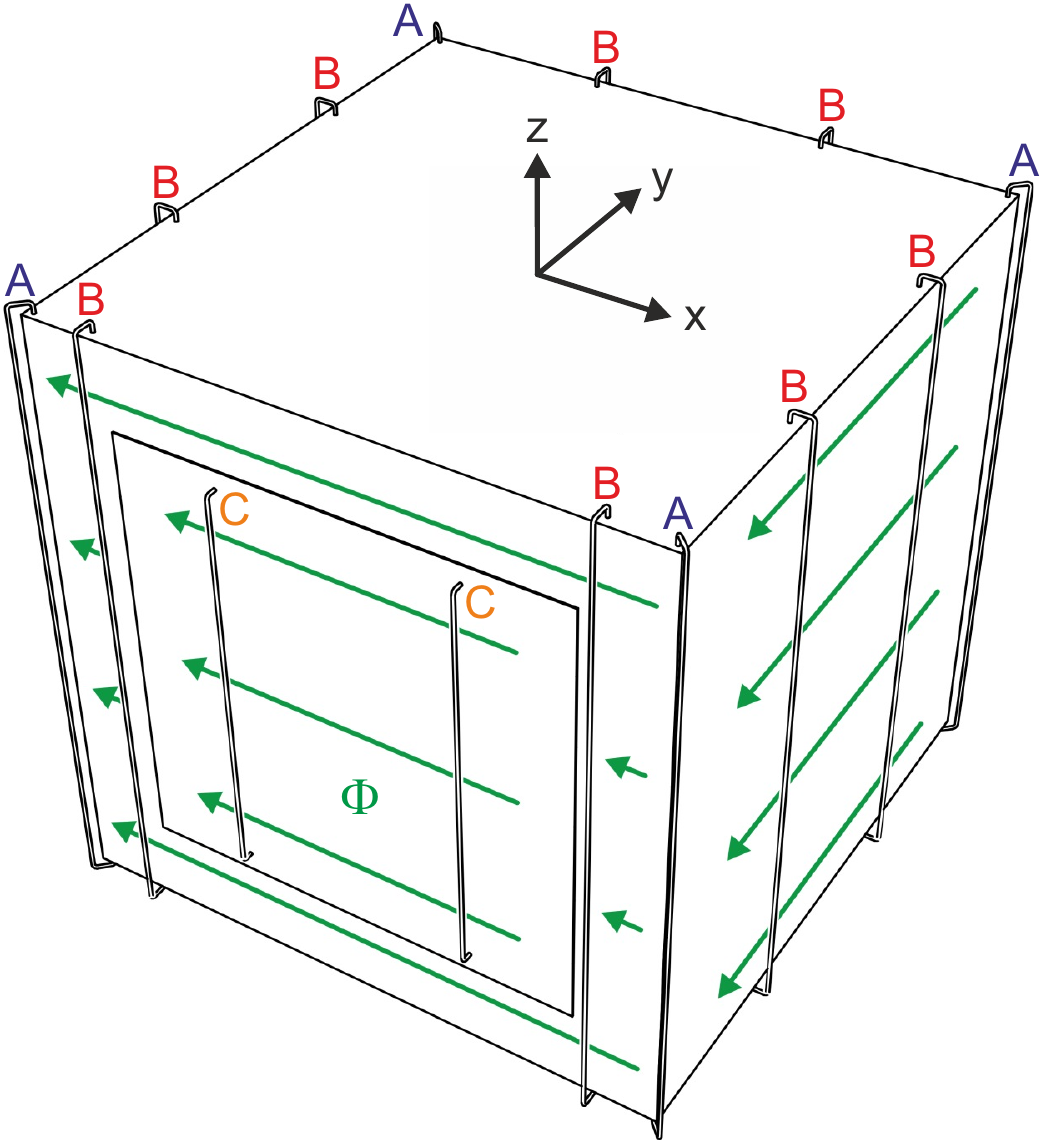}
\caption[]{
Arrangement of the equilibration coils in the $z$-direction
on one
%MUMETALL$^{{\tiny \textregistered}}$\
MSR layer
drawn as a cube box.
Label A: corner coils on all layers;
Label B: additional coils only on layer 6;
Label C: additional smaller coils only on the
layer 6 door.
The green arrows indicate the direction of the
magnetic flux $\Phi$
produced by a current through the indicated coils.
}
\label{fig:Degaussing-Coils}
\end{figure}

In order to minimize the remanent field in
the inner chamber, all MSR walls need to be
demagnetized~\cite{Thiel2007a},
or more precisely `equilibrated'~\cite{Voigt2013}
to achieve
the energetically most favorable state.
%{\bf thiel falsch dynamic modelling sollte man hier zitieren
%- allard cite }
This process is also sometimes colloquially referred to as `degaussing'.
Therefore, four coils per spatial direction
are installed with cables along every edge of every wall
of layer 1 to 6 individually,
as sketched in Fig.~\ref{fig:Degaussing-Coils},
similar to what is shown in Fig.~2 of Ref.~\cite{Voigt2013},
thus allowing the driving of a
magnetic flux independently in the three spatial dimensions.
Such an arrangement was first used
in Ref.~\cite{Voigt2013} for the `ZUSE' chamber
at PTB Berlin and was also used in Ref.~\cite{Altarev2015b}.
Here,
layer 6 has additional coils distributed
over the width of the walls and the door
to further improve the equilibration procedure
for the innermost layer~\cite{Sun2021}.

A reproducible and good equilibration result is obtained with
a sequential equilibration procedure
%
%A good equilibration is achieved
%with a procedure
driving an oscillating magnetic flux,
with the amplitude first increasing then
being slowly ramped down to zero.
%
%driving an increasing and then
%decreasing magnetic flux in each
%spatial direction of each layer.
%
All six layers are subsequently equilibrated
starting at the outermost layer.
In the initial characterization
measurements
a 5\,Hz sinusoidal signal
was used
to drive the current.
A standard equilibration procedure took
about 5 hours
and was repeated after every opening
of the MSR doors.
A more detailed description of the final
optimized equilibration
procedure will be part of a forthcoming publication.
%

%%%%%%%%%%%%%%%%%%%%%%%%%%%%%%%%%%%%%%%%%%%%%%%%%%%%%%%%%%%%%%%%%%%%%%%%%%%%%%%

\section{Remanent magnetic field}
\label{Remanent field}

\subsection{Measurement procedure}

\begin{figure}[h]
\centering
\includegraphics[width=0.9\columnwidth]{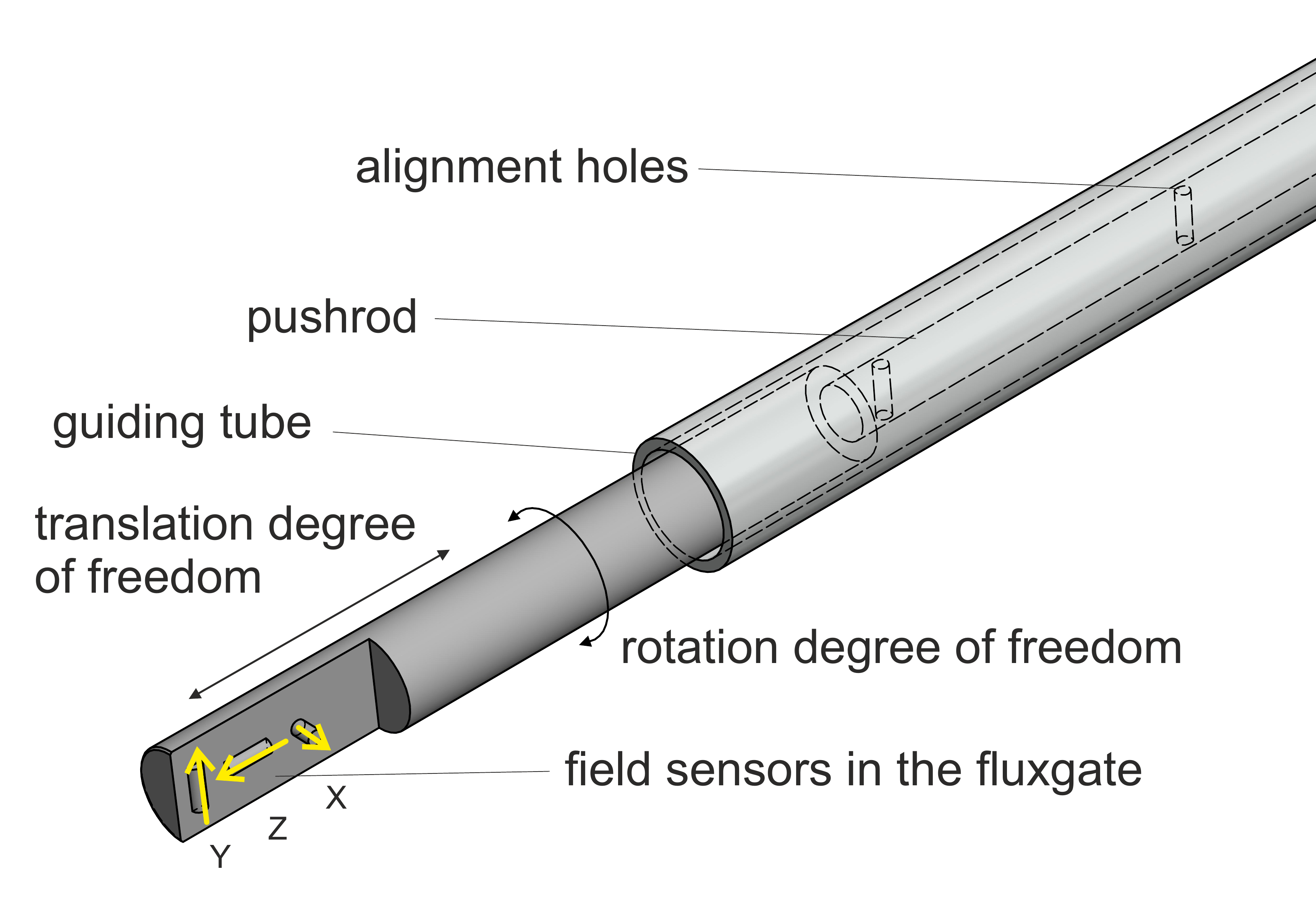}
\caption[]{
Scheme of the fluxgate sensor in the guiding tube.
The rotation degree of freedom is used to
determine the DC-offset for absolute field measurements for
the transverse sensors (x and y).
A pin through the holes in the pushrod was used to
fix both the rotation and translation.
The directions of the x, y and z component sensors are indicated
by the arrows.
% degrees of freedom.
}
\label{fig:fieldmeasurementprocedure}
\end{figure}

For this investigation the magnetic field in the inner chamber was measured with a
low-noise Bartington MAG03 three-axis
fluxgate~\footnote{Bartington Instruments Ltd,
Thorney Leys Park, Witney OX28 4GE, United Kingdom, Bartington.com}
located in a plexiglas tube installed between opposite openings
in the MSR walls.
Position scans were recorded by sliding the fluxgate along the axes
of this tube using a pushrod.
The rod has pin holes every 100\,mm that were
used to reproducibly fix the position along the tube as well
as the rotation of the
fluxgate around the axis of the tube.
The front part of this setup is depicted in
Fig.~\ref{fig:fieldmeasurementprocedure}.
The accessible measurement positions 
range from -60\,cm to +140\,cm
relative
to the center of the chamber.
At 140\,cm the fluxgate sensors are as close as 7\,cm to the
ULTRAVAC$^{{\tiny \textregistered}}$\ surface
of the innermost shielding layer.

A typical measurement consisted of integrating the
sensor signals for 3\,s and then rotating the fluxgate by 90 degrees.
This procedure was repeated until all four orientations
(0, 90, 180, 270 degrees)
of the fluxgate were recorded before proceeding to
the next
position along the tube.
The rotation allowed the
compensation of the
sensor offsets in the two transverse directions since
the contribution of the local magnetic field to the
sensor reading must invert when the sensors are rotated by 180 degrees.

When scanning in the vertical direction, the tube could be installed from the outside
so that the MSR doors did not have to be opened between measurements and the
equilibration procedure did not have to be repeated.
This means the magnetic configuration of the MSR was unchanged
except for possible relaxation processes in the wall material.
%MUMETALL$^{{\tiny \textregistered}}$.
%
All other measurements were performed after an equilibration of all shield layers.

\subsection{Results}

\begin{figure}[h]
\centering
\includegraphics[width=\columnwidth]{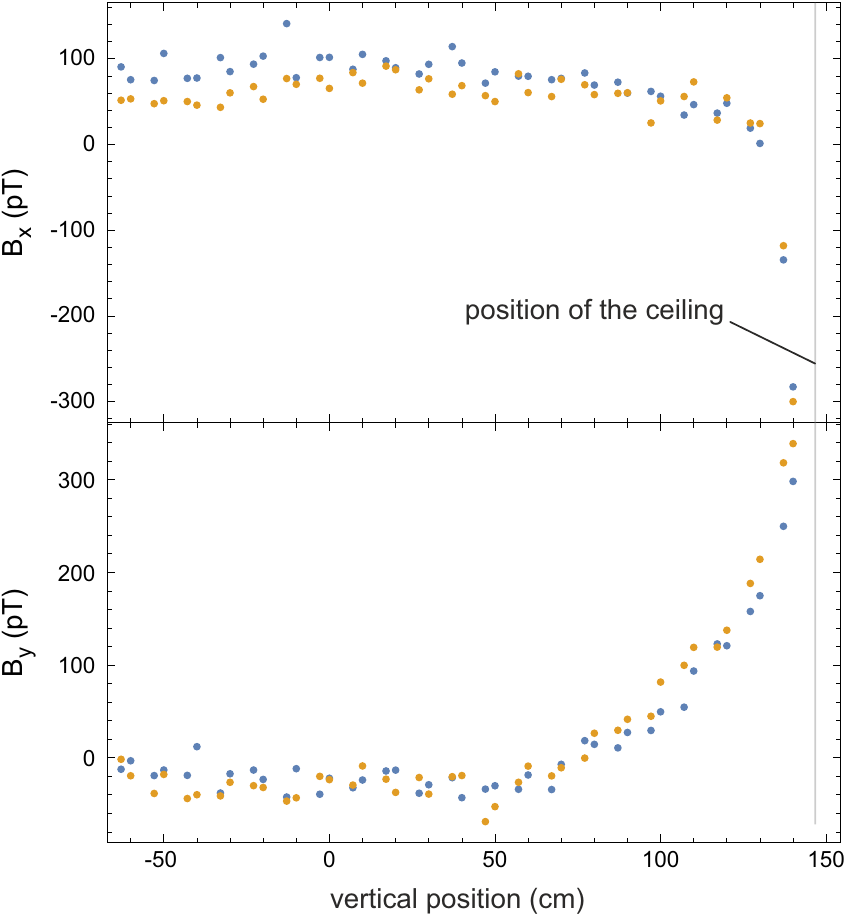}
\caption[]{
Results from two vertical scans of the remanent magnetic field inside
the MSR along the same axis.
The recordings were taken four days apart
with the blue dots indicating the first measurement.
The magnetic configuration of the MSR was not changed
during this time.
%
%Specifically the MSR was not opened and no equilibration was performed.
}
\label{fig:B-repeatability}
\end{figure}

\begin{figure}[h]
\centering
\includegraphics[width=\columnwidth]{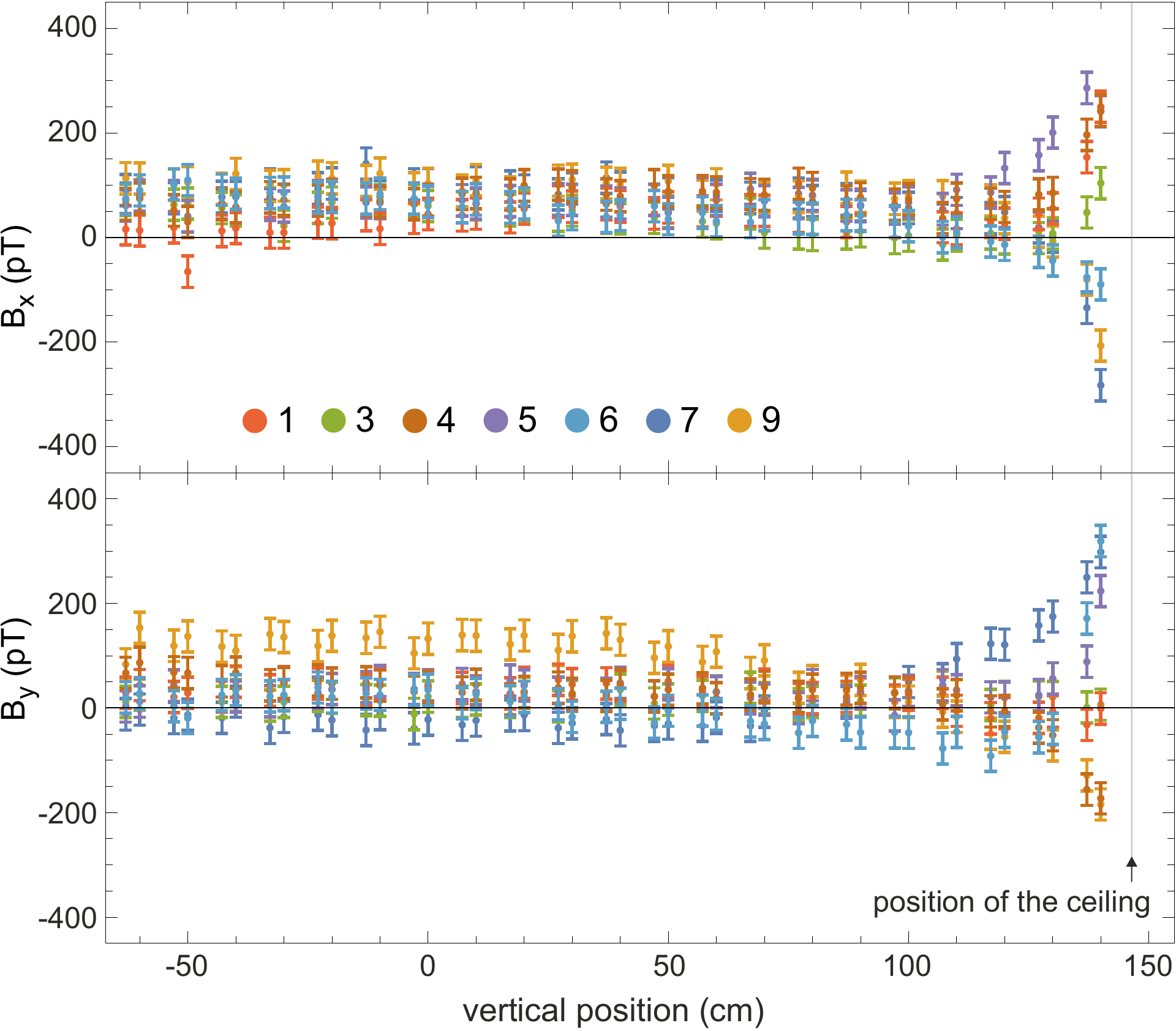}
\caption[Bxy]{
All B$_x$ and B$_y$ field values measured in the different vertical scans.
The remanent field increases to values of
about 300\,pT, only when approaching the
ULTRAVAC$^{{\tiny \textregistered}}$\ wall of the
inner chamber.
The colored points correspond to the measurements
at the position of the opening in the roof
as indicated by the number next to the colored filled circle and
with positions depicted in Fig.~\ref{fig:openings-floor}.
}
\label{fig:Bxy}
 \end{figure}

\begin{figure}[h]
\centering
\includegraphics[width=\columnwidth]{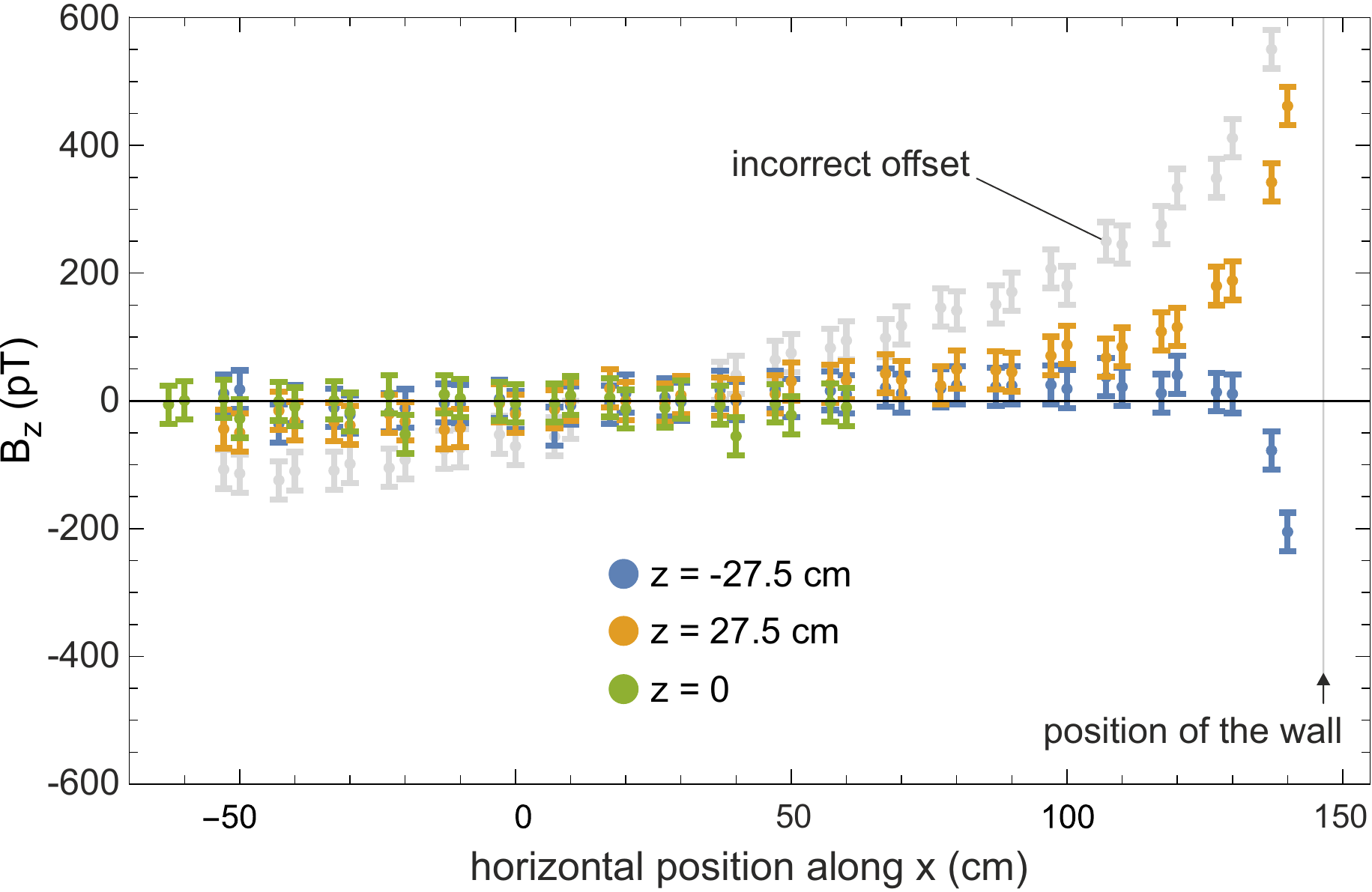}
\caption[Bz]{
B$_z$ field values measured in a horizontal scan along $x$
at the position of the
three large openings shown in the diagram of
Fig.~\ref{fig:MSR-intermediate-space},
at the center of the the MSR (z=0),
and above (27.5\,cm) and below (-27.5\,cm) the center.
%
%One can see the rise of
The remanent field increases to values of
about 500\,pT only when approaching the ULTRAVAC$^{{\tiny \textregistered}}$\ wall of the
inner chamber.
One measurement series depicted by the
gray dots display significantly larger remanent field values.
This was found later being due to an equilibration
procedure with an incorrect offset.
The gray color shows the measurement with incorrect
offset of the equilibration procedure.
The gray line indicates the position of the wall, i.e. layer 6.
}
\label{fig:Bz}
\end{figure}

\begin{figure}[h]
\centering
\includegraphics[width=\columnwidth]{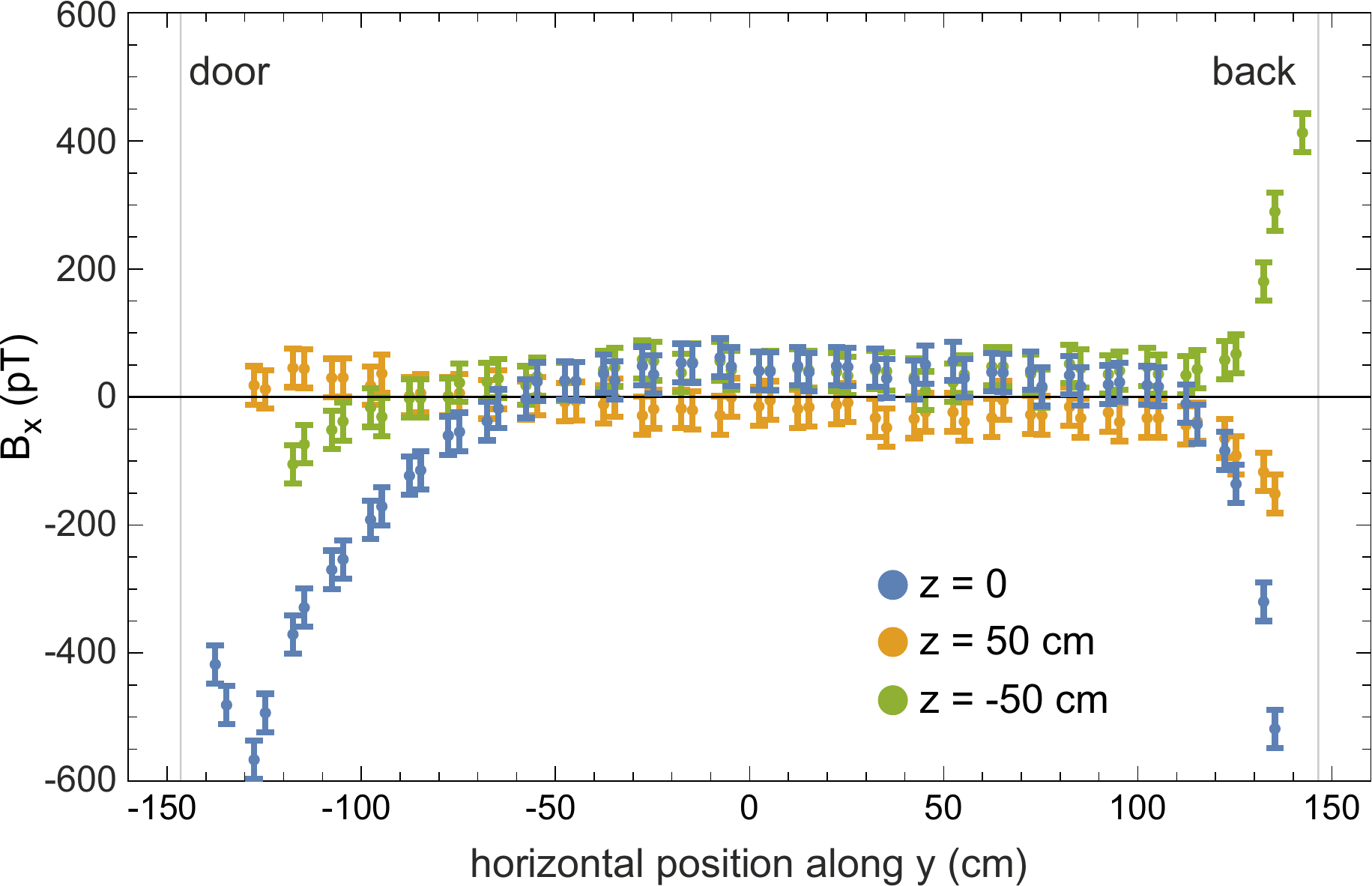}
\caption[]{
B$_x$ field values measured in a scan along the y axis
at the position of
three openings in the door
at the center (z=0),
and above (50\,cm) and below (-50\,cm) the center.
The gray lines indicate
the position of the door and the back wall.
}
\label{fig:Bxv}
\end{figure}

%All measured values for B$_x$ and B$_y$ shown in Fig.~\ref{fig:Bxy}
%were taken during vertical scans four days apart,
%and show the results after a single equilibration procedure.
%
In order to assess the repeatability of the magnetic field measurements
we repeated one vertical scan after four days.
Figure~\ref{fig:B-repeatability} compares the offset-corrected measurements from both scans.
The root mean square of the differences
between the two measurements
are 21\,pT and 24\,pT for the x- and y-directions, respectively.
The total deviation is close to the expected statistical uncertainty but shows also
a small systematic component, especially in $B_x$, where the mean difference
between all points of the scans amounts to 18\,pT.
Combining these two deviations, we conservatively estimate the total measurement error
to be 30\,pT which is reflected by the error bars shown in
Fig.~\ref{fig:Bxy}, \ref{fig:Bz} and \ref{fig:Bxv}.
%the following figures.

All measured values for B$_x$ and B$_y$
taken during vertical scans
in different positions
are shown in Fig.~\ref{fig:Bxy}
%were  four days apart,
and show the results after a single equilibration procedure.
Figure~\ref{fig:Bz}
shows the B$_z$ field component, which was measured in a horizontal scan
%The measurements shown in Fig.~\ref{fig:Bz}
%present the residual field
after equilibration performed on different days
since the doors had
to be opened in order to install the tube for the fluxgate.

The largest deviation from ideal behavior was found in a horizontal scan along the y-direction.
The corresponding measurements of B$_x$ are shown in Fig.~\ref{fig:Bxv}.
Here the mechanical scan range was increased to reach
from wall to wall.
%The measurements of the blue trace show that the door seems to have a certain influence on the field homogeneity in the inner chamber.
The measurements show the expected effect
that repeating an equilibration leads to
the strongest magnetic field uncertainties
close to the wall and especially close to the door.

Already with a non-optimized equilibration procedure
we find a large volume,
ranging from
-76\,cm to 76\,cm in x and y,
and -60\,cm to 140\,cm in z,
in which
all measured field values for B$_x$, B$_y$, and B$_z$
are below 150\,pT,
originally specified to be below 500\,pT.
Positions at lower x, y, and z values could not
be measured with the described setup.
%%%%
The gradients in the central 1\,m$^3$
were significantly smaller than originally
specified (300\,pT/m, see Section~\ref{Sec:Specification}).
In this volume the gradients are within the statistical
uncertainty of the
fluxgate sensors used,
which is
estimated to correspond to 60\,pT/m at 1$\sigma$ confidence level.

\begin{figure}[h]
\centering
\includegraphics[width=\columnwidth]{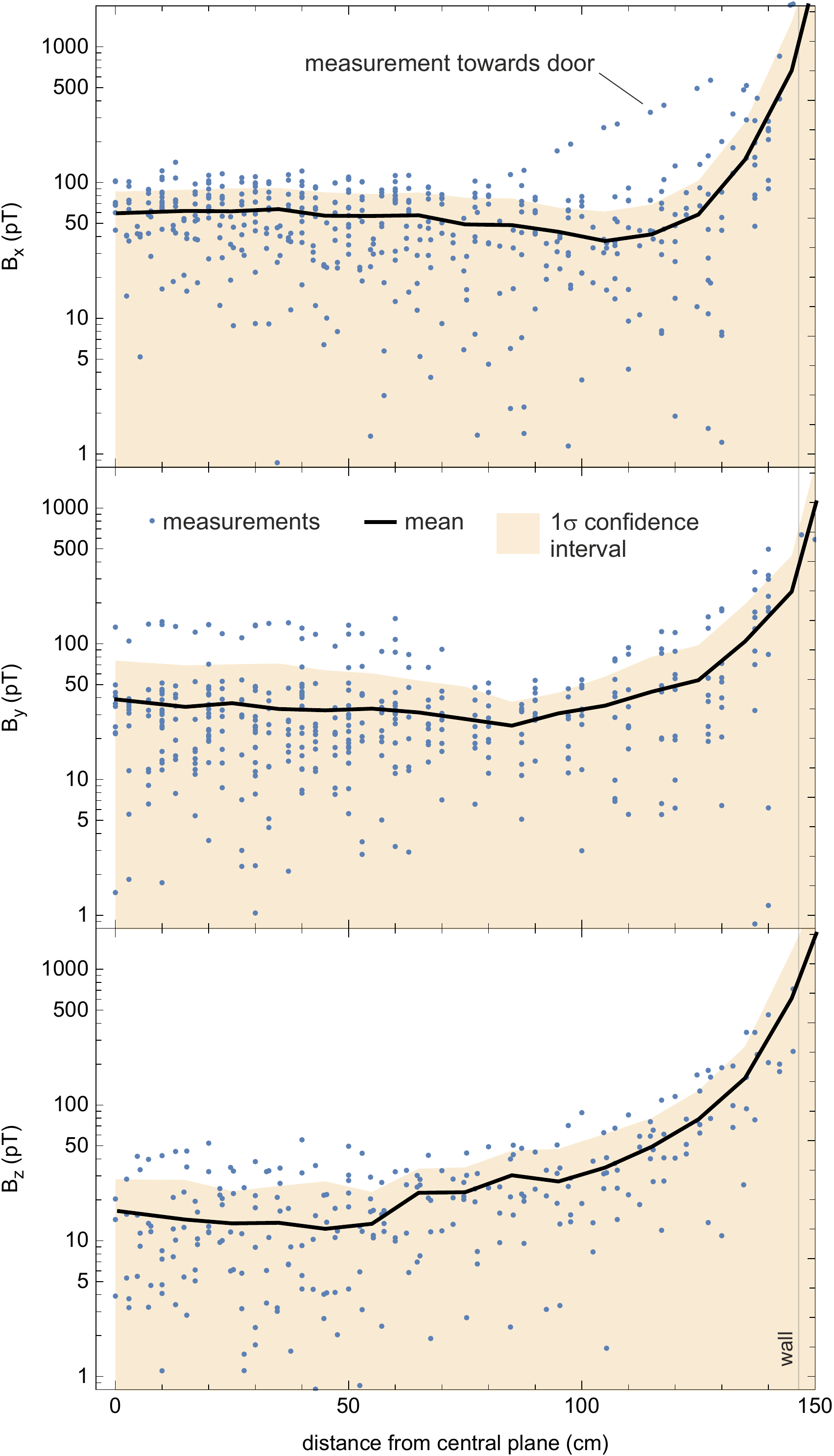}
\caption[]{
Summary of all measured scans
of the three field components B$_x$, B$_y$ and B$_z$
performed after equilibration of the MSR.
The magnitude of the magnetic field components
is shown as a function of distance to
a central plane perpendicular to the scan direction.
All points in the central cubic meter thus fall into the
region with distance smaller than 50\,cm.
The black lines gives the mean values with the
1$\sigma$ uncertainties displayed as the shaded area.
}
\label{fig:remanentb}
\end{figure}

The remanent magnetic field measurements along
the three spatial directions
are summarized in
Fig.~\ref{fig:remanentb}
%with respect to
relative to
their distance from the center of the inner chamber.
This distance is
computed relative to a plane through the center and
perpendicular to the scan direction.
Hence, all measured points in the central 1\,m$^3$ volume are
at a distance <50\,cm.
The ULTRAVAC$^{{\tiny \textregistered}}$\ material
of the layer 6 wall is at a distance of 146.5\,cm.
One can see that all $1\sigma$ confidence intervals
in the central 8\,m$^3$
(positions <100\,cm)
are below 100\,pT.
For the z-component of the magnetic field, which is most 
important for n2EDM, also the maximum deviation is below 
100\,pT in the central 8\,m$^3$.
Only when approaching the
wall,
the remanent field values slowly increase
to approximately 500\,pT at
a distance of about 4\,cm from the ULTRAVAC$^{{\tiny \textregistered}}$.

%%%%%%%%%%%%%%%%%%%%%%%%%%%%%%%%%%%%%%%%%%%%%%%%%%%%%%%%%%%%%%%%%%%

\section{Summary}
\label{Summary}

We constructed a unique
magnetically shielded room with excellent performance,
providing 25\,m$^3$ of usable shielded volume
for the n2EDM apparatus,
which will search for the neutron
electric dipole moment with
a baseline sensitivity of 10$^{-27}$\,e\,cm.
This MSR provides
the largest ultra-low magnetic field environment
in the world
despite the numerous openings
%with many large openings
allowing for access and
throughgoing connections.

The excellent magnetic performance is achieved using
five MUMETALL$^{{\tiny \textregistered}}$\ layers,
one ULTRAVAC$^{{\tiny \textregistered}}$\ layer,
and one aluminum layer.
A quasi-static shielding factor at 0.01\,Hz of
$\sim$100,000 was measured in all three spatial directions.
The shielding factor rapidly raises
with frequency and already reaches
10$^8$ for frequencies above 3\,Hz.

After applying the equilibration procedure
the MSR provides  an exceptionally low magnetic 
field environment across a large volume.
As the other two magnetic field components, 
in particular
the most important field component $B_z$ (vertical)
shows remanent magnetic field values below 100\,pT 
in the central 8\,m$^3$.

\section{Acknowledgements}

We especially acknowledge the dedicated work
of the group for magnetically shielded rooms
of the company
VAC - Vacuumschmelze, Hanau; namely of
Lela Bauer,
Markus Hein
and
Michael W\"ust.
Excellent support throughout the planning
and construction by Michael Meier was essential.
Thomas Stapf supported the design
and installation.
Technical support by Luke Noorda
is acknowledged.
We appreciate the help of the
many PSI support groups
contributing to the installation.

Financial support by the Swiss National Science Foundation
R`EQUIP 139140, % MSR
by PSI,
and
by the Emil-Berthele-Fonds was crucial for this work.

The material scans inside BMSR-2 were supported by the Core Facility
`Metrology of Ultra-Low Magnetic Fields' at Physikalisch-Technische Bundesanstalt
which received funding from the Deutsche Forschungsgemeinschaft –
DFG (funding codes: DFG KO 5321/3-1 and TR 408/11-1).

Support by the Swiss National Science Foundation Projects
200020-188700 (PSI), %Bernhard neu 2020++
200020-163413 (PSI), %Bernhard 2015-2019
200011-178951 (PSI), % GEza new
172626 (PSI), % Georg magnetometer 2017 Duarte
169596 (PSI), % PSW - PinJung
%200020-137664 (PSI),  % Bernhard UCN 2011-2013
%200021-117696 (PSI),  % Klaus PSI EDM  2008-11
%200020-144473 (PSI),  % GEza 2012 - 2014
%200021-126562 (PSI),  % GEza 2009 - 2012
200021-181996 (Bern),  % Piegsa 2019 - 2022
%200020-172639 (ETH), %Klaus alt
200441 (ETH),  %Klaus new
%R`EQUIP 139140, % MSR
%177008 % PSW HV supply
and FLARE 20FL21-186179, % Flare 1
and 20FL20-201473 % Flare 2 / 2021
is gratefully acknowledged.
The LPC Caen and the LPSC Grenoble acknowledge the support of the
French Agence Nationale de la Recherche (ANR) under
reference ANR-14-CE33-0007 and the ERC project 716651-NEDM.
University
of Bern acknowledges the support via the European Research Council under
the ERC Grant Agreement No. 715031-Beam-EDM.
The Polish collaborators wish to acknowledge support from the
National Science Center, Poland, under grant
%No. 2015/18/M/ST2/00056,
%No. 2016/23/D/ST2/00715,
No. 2018/30/M/ST2/00319, and No. 2020/37/B/ST2/02349.
Support by the Cluster of Excellence
`Precision Physics, Fundamental Interactions, and Structure of Matter'
(PRISMA+ EXC 2118/1) funded by the German Research Foundation (DFG)
within the German Excellence Strategy
(Project ID 39083149) is acknowledged.
Collaborators at the University of Sussex wish to acknowledge
support from the School of Mathematical and Physical Sciences,
as well as from the STFC under grant ST/S000798/1.
This work was partly supported by the Fund for Scientific
Research Flanders (FWO), and Project GOA/2010/10 of the KU Leuven.
Researchers from the University of Belgrade
acknowledge institutional funding provided by
the Institute of Physics Belgrade through a
grant by the Ministry of Education, Science
and Technological Development of the Republic of Serbia.

% If in two-column mode, this environment will change to single-column format so that long equations can be displayed.
% Use only when necessary.
%\begin{widetext}
%$$\mbox{put long equation here}$$
%\end{widetext}

\section{Data availability}

The datasets generated and/or analyzed during the current study are not publicly 
available but are available from the corresponding authors on reasonable request.

\section{References}

%\bibliography{TDR-references,nedm-references,BSM-references,UCN-references,SM-references,MSR-references}
\bibliography{MSR.bbl}

\end{document}